\documentstyle[epsfig,aps]{revtex}

\newcommand{\ccap}[2]{\caption[#1]{#2}\label{#1}\vspace{0.2cm}}
\begin{document}
\draft
\title{ Anomalous $Wtb$ Coupling in 
$ep$ Collision}
\mediumtext
\author{S. Ata\u{g}\footnotemark,  O. \c{C}ak\i r and B. Dile\c{c}} 
\address{Ankara University Department of Physics \\
 Faculty of Sciences, 06100 Tandogan, Ankara, Turkey}
\footnotetext{Corresponding author: Fax:+90 312 2232395, 
e-mail: atag@science.ankara.edu.tr}
\maketitle

\begin{abstract}
The potential of $ep$ collision
to prospect for anomalous 
$Wtb$ vertex is discussed from the single 
top quark production  process 
$ep\rightarrow t\bar{\nu}+X$ for TESLA+HERAp 
and CLIC+LHC energies.
Sensitivities to anomalous couplings 
$F_{2L}$ and $F_{2R}$, 
in the case of CLIC+LHC, are shown to be 
comparable with LHC.
\end{abstract} 

\vskip 0.5cm

\section{Introduction}
After the observation of top quark with mass much heavier 
than the rest of the Standard Model(SM) fermions and gauge bosons, 
special attention to its couplings with gauge bosons has been 
drawn. Since the coupling $Wtb$ is responsible for all top 
quark  decays, it plays crucial role to help understand the 
nature of electroweak theory and "new physics" beyond. 
Therefore it is important to study  $Wtb$ vertex and measure
the coupling parameters with high precision. 
Single top quark production processes provide unique possibility
to search for this vertex due to their direct proportionality to  
the $Wtb$ coupling. Deviations from the SM expectation in 
$Wtb$ vertex would be a possible signal for the new physics 
beyond SM. Several collider experiment potentials have been 
examined to search for these coupling parameters through single 
top production. Single top cross section for the process 
$e^{+}e^{-}\rightarrow Wtb$ has been discussed below and 
above  $t\bar{t}$ threshold \cite{mele,jikia} and  for the 
process $e^{+}e^{-}\rightarrow e\bar{\nu}tb$ at CERN 
$e^{+}e^{-}$ collider LEP2 \cite{hagiwara} and linear 
$e^{+}e^{-}$ collider(LC) \cite{boos,bien} energies. 
Investigations for $Wtb$ vertex have been done 
at $\gamma e$ mode of LC \cite{pukhov},
Fermilab $p\bar{p}$ collider  Tevatron and CERN $pp$ collider 
LHC \cite{dicus,dudko}. 

Additional  option of linear $e^{+}e^{-}$ collider  would be 
an $ep$ collider when LC is constructed on the same 
base as the proton ring. Linear collider design TESLA at DESY 
is the one that can be converted into TESLA+HERAp  $ep$ collider 
\cite{thera}. Similar option would be considered for CLIC+LHC
at CERN. Estimations about the main parameters of these 
collider modes are shown in Table \ref{tab1} for two different 
design values of linear electron beam energies. 

In this paper, the potential of future high energy ep colliders
to investigate anomalous $Wtb$ couplings will be discussed.  

\section{ Lagrangian and Cross Sections}
In the model independent effective lagrangian approach
\cite{bmuller,renard,yang,ladinsky}  there
are seven anomalous CP conserving operators of dimension six
which contribute $Wtb$ vertex \cite{yang} . This effective
lagrangian contains four independent couplings whose
explicit forms are given in ref. \cite{yang}.
Effects of all seven operators will not be investigated
here. We only consider following   couplings
to reveal the potential of $ep$ collision

\begin{eqnarray}
L&&={{g_{w}}\over {\sqrt{2}}}
[W_{\mu}\bar{t}(\gamma^{\mu}F_{1L}P_{-}
+\gamma^{\mu}F_{1R}P_{+})b
-{{1}\over {2m_{w}}}
W_{\mu\nu} \bar{t}\sigma^{\mu\nu}
(F_{2L}P_{-}+F_{2R}P_{+})b] + h.c.
\end{eqnarray}
where 

\begin{eqnarray}
W_{\mu\nu}=D_{\mu}W_{\nu}-D_{\nu}W_{\mu} \;\;,\;\;
D_{\mu}=\partial_{\mu}-ieA_{\mu} \nonumber \\
P_{\mp}={1\over 2}(1\mp\gamma_{5}),  \;\;,\;\;
\sigma^{\mu\nu}={i\over 2} (\gamma^{\mu}\gamma^{\nu}
-\gamma^{\nu}\gamma^{\mu})
\end{eqnarray}
In the SM, the (V-A) coupling $F_{1L}$ corresponds 
Cabibbo-Kobayashi-Maskawa (CKM) 
matrix element $V_{tb}$ which is very close to unity and 
$F_{1R}$, $F_{2L}$ and $F_{2R}$ are equal to zero. The 
(V+A) coupling $F_{1R}$ is severely bounded by the CLEO 
$b\rightarrow s\gamma$ data \cite{cleo} at a level such 
that it will be out of reach at expected future colliders.
Therefore we set $F_{1L}=0.999$ and $F_{1R}=0$ as required 
by present data \cite{pdg}. The magnetic type anomalous 
couplings are related to the coeficients $C_{tW\Phi}$ and 
$C_{bW\Phi}$ \cite{yang} in the general effective 
lagrangian by 

\begin{eqnarray}
F_{2L}={{C_{tW\Phi}\sqrt{2}v m_{w} }\over{\Lambda^{2}g}}
\;\;\;\; 
F_{2R}={{C_{bW\Phi}\sqrt{2}v m_{w} }\over{\Lambda^{2}g}}
\end{eqnarray}
where $\Lambda$ is the scale of new physics. Natural 
values of the couplings $F_{2L(R)}$  are in the region
\cite{peccei}  of 
\begin{eqnarray}
{\sqrt{m_{b}m_{t}}\over {v}}\sim 0.1
\end{eqnarray}
and do not exceed unitarity violation bounds for 
$|F_{2L(R)}|\sim 0.6$ \cite{renard}.

In $ep$ collision there are two subprocesses contributing 
to single top production $eb\rightarrow t\bar{\nu}$ and 
W-gluon fusion process $eg\rightarrow t\bar{b}\bar{\nu}$.
Differential cross section for the subprocess 
$eb\rightarrow t\bar{\nu}$ in terms of Mandelstam 
invariants $\hat{s}$ and $\hat{t}$ has been obtained 
as given below 

\begin{eqnarray}
{{d\hat{\sigma}}\over{d\hat{t}}}=&&{{\pi\alpha^{2}}
\over{4\sin^{4}\theta_{W}\hat{s}^{2}m_{w}^{2}(\hat{t}
-m_{w}^{2})^{2} }}\times \nonumber \\
&&[F_{2L}^{2}\hat{t}
(-(\hat{s}-m_{t}^{2})^{2} +\hat{t}(m_{t}^{2}-\hat{s})) 
+2F_{2L} |V_{bt}|m_{t}m_{w}\hat{t}(\hat{s}
+\hat{t}-m_{t}^{2}) \nonumber \\
&&-F_{2R}^{2}\hat{s}\hat{t}(\hat{s}+\hat{t})
+|V_{bt}|^{2}m_{w}^{2}(-m_{t}^{2}(\hat{s}+\hat{t})
+(\hat{s}+\hat{t})^{2})] 
\end{eqnarray}
where $m_{t}$, $m_{w}$, $V_{bt}$  and $\theta_{W}$ are top 
quark mass, W boson mass, CKM matrix element and Weinberg angle
respectively.

If the  second process  $eg\rightarrow t\bar{b}\bar{\nu}$ 
is combined with $eb\rightarrow t\bar{\nu}$ one should
avoid double counting. The two subprocesses have a region 
of overlap when $g\rightarrow b\bar{b}$ in the W-gluon 
fusion is nearly collinear (close to on shell) 
so that $b$ quark interacting with 
W boson is the same as the initial $b$ sea quark in the 
first subprocess $eb\rightarrow t\bar{\nu}$. In order 
to take care of this double counting we use the method 
proposed in ref. \cite{collins}. So, combined cross section 
becomes 
\begin{eqnarray}
[\sigma(eb \to t\bar{\nu})+ \sigma(eg \to t\bar{b}\bar{\nu})
-\sigma(g\to b\bar{b}*eb \to t\bar{\nu})]
\end{eqnarray}    
where the subtracted term is the gluon splitting piece of 
the cross section for $eg\to t\bar{b}\bar{\nu}$ and they 
are all integrated cross sections over the parton 
distributions.  For the distribution function  of 
$b$ quark inside the gluon (splitting part) leading order 
approximation will be used defined below

\begin{eqnarray}
\sigma_{ex}(ep\rightarrow t\bar{\nu}+X)=\int_{m_{t}^{2}/s}^{1}
\hat{\sigma}(xs)[f_{b/p}(x,Q^{2})-f_{b/p}^{LO}(x,Q^{2})]dx
\end{eqnarray}
where

\begin{eqnarray}
f_{b/p}^{LO}(x,Q^{2})={{\alpha_{s}(Q^{2})}\over{2\pi}}
\int_{x}^{1}{{d\xi}\over{\xi}}P_{b/g}({{x}\over{\xi}})
f_{g/p}(\xi,Q^{2})\ln{{Q^{2}}\over{m_{b}^{2}}}
\end{eqnarray}  
with

\begin{eqnarray}
P_{b/g}(z)={{1}\over{2}}[z^{2}+(1-z)^{2}]
\end{eqnarray} 
Here, $\alpha_{s}(Q^{2})$ is the energy dependent QCD  coupling 
whose expression is given in ref. \cite{pdg}. 
$\sigma_{ex}$ is the integrated 
total cross section over the parton distributions 
for the subprocess $eb \to t\bar{\nu}$ 
where gluon splitting piece is 
subtracted. The cross section for the subprocess 
$eg \to t\bar{b}\bar{\nu}$ is still to be added to 
$\sigma_{ex}$ for the combined cross section. 
QCD scale, $Q^{2}$, dependence of these cross 
sections is presented 
in Table \ref{tab2} in the region $m_{w}/2<Q<2m_{t}$ for 
SM values of the couplings at $\sqrt{s}=1.6$ TeV.
It is seen that as $Q$ increases $\sigma_{eb}$ increases 
while $\sigma_{eg}$ decreases. Then, the combined cross section 
$\sigma$ is slightly changed by the variation of $Q^{2}$.   

Table \ref{tab3} and Table \ref{tab4} show the influence of the 
the anomalous couplings $F_{2L}$, $F_{2R}$ on the 
integrated total cross sections and subtraction 
at TESLA+HERAp and CLIC+LHC energies for $Q^{2}=m_{w}^{2}$
and $m_{t}=175$ GeV.
Parton distribution functions of Martin, Robert and Stirling 
(MRS A) \cite{martin} have been used  
and total cross section for second process
$eg\rightarrow t\bar{b}\bar{\nu}$ has been calculated
using CompHEP package \cite{chep}.
As seen from tables subtracted terms are not negligible and 
the combined cross sections are slightly larger than the 
cross section of the first process  $\sigma_{eb}$ 
by a factor 1.1 for TESLA+HERAp and by a factor 1.2
for  CLIC+LHC. From here on, we will consider only the 
first process because of its simplicity
to investigate potential sensitivity of $ep$ collision 
to anomalous $Wtb$ vertex. Integrated total cross section 
as a function of center of mass energy $\sqrt{s}$ 
of $ep$ system is shown 
in Fig. \ref{fig1} for SM and $F_{2L}=-0.2$. From Tables 
\ref{tab3}, \ref{tab4} and Fig. \ref{fig1}, it is clear 
that total cross sections in $ep$ collision  are much 
larger than the case of $\gamma e$ mode \cite{pukhov}
of $e^{+}e^{-}$ collider  and Tevatron \cite{dudko}. 
Furthermore, cross sections from CLIC+LHC are a few 
times larger than  those of LHC \cite{dudko} as expected. 
Fig. \ref{fig2} shows the cross sections as functions of
the anomalous couplings $F_{2L}$ and $F_{2R}$ 
for TESLA+HERAp and CLIC+LHC energies. A common feature 
of each figure is that  deviation from the SM 
increases with increasing energy and  cross sections grow 
as deviation gets large.  

The influence of the deviation of anomalous couplings 
from the SM on the shape of the angular and 
$p_{T}$ distribution of  top quark 
is also important. Integrated differential cross sections 
as functions of the  angle between  top quark and incoming 
proton direction in the center of mass frame of 
$t\bar{\nu}$ are plotted in Fig. \ref{fig3}. 

In order to get $p_{T}$ distribution of top quark we 
use following standard procedure:

\begin{eqnarray}
{{d\sigma}\over{dp_{T}}}=2p_{T}\int_{y^{-}}^{y^{+}}
{{dy}\over{|s-2E_{p}m_{T}e^{y}|}}
f_{q/p}(x,Q^{2})\hat{s}{{d\hat{\sigma}}\over{d\hat{t}}}
\end{eqnarray}
where $y$ is the rapidity of the top quark whose 
lower and upper limits and momentum fraction $x$ of 
the struck quark in the proton are given by
\begin{eqnarray}
y^{\mp}=&&\ln{[{{s+m_{t}^{2}}\over{4E_{p}m_{T}}}
\mp\{({{s+m_{t}^{2}}\over{4E_{p}m_{T}}})^{2}
-{{E_{e}}\over{E_{p}}}\}^{1/2}]} \\
x=&&{{2E_{e}m_{T}e^{-y}-m_{t}^{2}}\over{s-2E_{p}m_{T}e^{y}}}.     
\end{eqnarray}
Here $E_{p}$ and $E_{e}$ are proton and electron beam 
energies and $\hat{s}$, $\hat{t}$ and $m_{T}$ 
are defined as follows
\begin{eqnarray}
\hat{s}=xs, \;\;\;\; \hat{t}=m_{t}^{2}-2xE_{p}m_{T}e^{y},
\;\;\;\; m_{T}^{2}=m_{t}^{2}+p_{T}^{2}.
\end{eqnarray}
The behaviour of $p_{T}$ spectrum of top quark is 
shown in Fig. \ref{fig4} for two different energy region 
and for some anomalous couplings.
Clearly, angular and $p_{T}$ distributions of top quark 
lead to deviations from the SM expectation.

\section{Sensitivity to Anomalous Couplings}

We use simple $\chi^{2}$ criterion from angular 
distributions of top quark to estimate  sensitivity 
of $ep$ collision to anomalous $Wtb$ couplings 

\begin{eqnarray}
\chi^{2}=\sum_{i=bins}({{X_{i}-Y_{i}}\over{\Delta_{i}^{exp}}})^{2}
\end{eqnarray}
\begin{eqnarray}
X_{i}=\int_{z_{i}}^{z_{i+1}} {{d\sigma^{SM}}\over{dz}}dz ,
\;\;\;\;\;\;
Y_{i}=\int_{z_{i}}^{z_{i+1}} {{d\sigma^{NEW}}\over{dz}}dz
\end{eqnarray}

\begin{eqnarray}
\Delta_{i}^{exp}=X_{i}\sqrt{\delta_{stat}^{2}+\delta_{sys}^{2}}
\;,\;\;\;\;\;\;z=\cos{\theta}.
\end{eqnarray}

We have divided the range of $\cos{\theta}$  into 6 pieces 
for TESLA+HERAp and 10 pieces for CLIC+LHC 
and  have considered at least 20 events in each bin. 
The expected number of events in the i-th bin which is 
used in statistical error has been 
calculated considering the leptonic channel of W boson as 
the signal  
$N_{i}=\epsilon L_{int}\sigma_{i}BR(W\rightarrow \ell+\nu)$ 
where $\epsilon$ is the overall efficiency and $L_{int}$
is the integrated luminosity.  
The limits on the anomalous couplings 
$F_{2L}$ and $F_{2R}$ are provided 
in Table \ref{tab5} at TESLA+HERAp and Table \ref{tab6} at 
CLIC+LHC energies for the deviation from the SM values at 
95\% confidence level. 
Only one of the couplings is assumed 
to deviate from the SM at a time.   
With integrated luminosities in Table \ref{tab5} the potential 
sensitivities of TESLA+HERAp to both $F_{2L}$ and $F_{2R}$ 
are about $O(10^{-2})$ ($O(10^{-1})$ with 10\% systematic) 
in the case of higher energy option 
$\sqrt{s}=1.6$ TeV which improve the results obtained 
at Tevatron  \cite{dudko}. 
For possible CLIC+LHC energies from 
Table \ref{tab6}, sensitivities to $F_{2L}$ and $F_{2R}$ 
are about $O(10^{-2})$ ($O(10^{-1})$ with 10\% systematic).
For $\sqrt{s}=6.5$ TeV  region, CLIC+LHC will have 
higher potential to probe $F_{2L}$ and $F_{2R}$ than LHC 
\cite{dudko}. 
In order to compare ep colliders with $\gamma e $ mode of 
linear $e^{+}e^{-}$ collider \cite{pukhov}, 
hadronic channels should be 
included with more reduced  uncertainties. 

Systematic uncertainties from $V_{tb}$, $m_{t}$, 
parton distribution functions, QCD scales, 
luminosity measurement are also important for accurate 
results. However, at this stage it is difficult to give
a realistic estimate of systematics. Therefore, combined 
systematic errors of 0.05 and 0.10 are taken into account 
and $\epsilon=0.5$ overall efficiency is assumed.
For more precise results, further analysis
needs to be supplemented by observables such as the
distributions of the top decay products i.e.,
$ep\rightarrow \ell+b+\bar{\nu}+\nu_{\ell}+X$
with a more detailed knowledge of the experimental 
performances.


\begin{table}[bth]
\caption{ Main parameters of $ep$ colliders where 
linear electron beams are allowed to collide protons 
from the ring for two different design values of 
linear electron beam energies. Luminosity values reflects the 
orders only. 
\label{tab1}}
\begin{center}
\begin{tabular}{lll}
Colliders & $\sqrt{s_{ep}}$(TeV) & $L_{ep}(cm^{-2}s^{-1})$ \\
\hline
CLIC+LHC & 5 & $10^{32}$ \\
CLIC+LHC & 6.5  & $10^{32}$ \\
TESLA+HERAp & 1 & $10^{31}$ \\
TESLA+HERAp & 1.6 & $10^{31}$ \\
\end{tabular}
\end{center}
\end{table}

\begin{table}[bth]
\caption{$Q^{2}$ dependence of the 
cross sections $\sigma_{eb}$ from  the 
subprocesses $eb\rightarrow t\bar{\nu}$,
$\sigma_{eg}$ from $eg \rightarrow t\bar{b}\bar{\nu} $
and combined cross section $\sigma$ 
where on-shell b-quark
contribution was subtracted. 
Standard Model values of the couplings are considered 
only and cross sections are computed for $\sqrt{s}=1.6$ TeV.  
\label{tab2}}
\begin{center}
\begin{tabular}{llll}
 $Q $   & $\sigma_{eb}$ & $\sigma_{eg}$
 & $\sigma$  \\
\hline
$2m_{t}$ &3.78 &1.88 &3.28  \\
$m_{t}$ &3.55 &2.13 &3.38  \\
$m_{w}$&3.20 &2.53 &3.59 \\
$m_{w}/2$&2.79 &3.00 &3.88 \\
\end{tabular}
\end{center}
\end{table}

\begin{table}[bth]
\caption{Integrated total cross sections of the process
$ep\rightarrow t\bar{\nu}+X $ in pb for 
$\sqrt{s}=1.6$ TeV, TESLA+HERAp collider and for $Q=m_{w}$.
Cross sections $\sigma_{eb}$ and $\sigma_{eg}$ are contributions 
from the subprocesses $eb\rightarrow t\bar{\nu}$
and $eg \rightarrow t\bar{b}\bar{\nu} $. 
$\sigma$ is combined cross section where on-shell b-quark
contribution was subtracted from 
$eb\rightarrow t\bar{\nu}$ before combination 
to avoid double counting. 
\label{tab3}}
\begin{center}
\begin{tabular}{lllll}
 $F_{2L}$ &$F_{2R}$   & $\sigma_{eb}$ & $\sigma_{eg}$ 
 & $\sigma$  \\
\hline
0& 0 &3.20 &2.53 &3.52  \\
-0.1& 0 &3.33 &2.60 &3.63 \\
0.1& 0 &3.13 &2.51 &3.48 \\
0& 0.1 &3.28 &2.61 &3.63  \\
0& 0.2 &3.51 &2.88 & 3.97 \\
-0.1& 0.1 &3.41 &2.70 & 3.76 \\
\end{tabular}
\end{center}
\end{table}

\begin{table}[bth]
\caption{The same as Table \ref{tab2} but for 
$\sqrt{s}=5$ TeV, CLIC+LHC energy. 
\label{tab4}}
\begin{center}
\begin{tabular}{lllll}
 $F_{2L}$ &$F_{2R}$   & $\sigma_{eb}$ & $\sigma_{eg}$
 & $\sigma$  \\
\hline
0& 0 &29.0 &25.6 &33.6  \\
-0.1& 0 &30.2 &26.1 &34.4 \\
0.1& 0 &28.7 &25.6 &33.5 \\
0& 0.1 &29.7 &26.2 &34.4  \\
0& 0.2 &32.1 &29.6 & 38.5 \\
-0.1& 0.1 &30.9 &27.2 & 35.8 \\
\end{tabular}
\end{center}
\end{table}

\begin{table}
\caption{Sensitivity of TESLA+HERAp  collider to
anomalous $Wtb$ couplings at 95\% C.L. 
Only one of the couplings is assumed to deviate from the SM
at a time.
\label{tab5}}
\begin{center}
\begin{tabular}{lllcc}
$\sqrt{s_{ep}}$(TeV).&$\int{L}dt(fb^{-1})$ &$\delta^{sys}$ 
&$F_{2L}$ & $F_{2R}$ \\
\hline
 1 & 20 & 0 &  -0.075, 0.558 &  -0.100, 0.100 \\ 	
 1 & 20 & 0.05     &-0.084, 0.603 & -0.105, 0.105   \\
 1 & 20 & 0.10  &-0.099, 0.636 & -0.112, 0.112   \\
 1.6 & 20 & 0 &  -0.040, 0.051  & -0.072, 0.072 \\
 1.6 & 20 & 0.05     &-0.051, 0.069 & -0.079, 0.079   \\
 1.6 & 20 & 0.10  &-0.066, 0.465 & -0.088, 0.088   \\
\end{tabular}
\end{center}
\end{table}
 
\begin{table}
\caption{Sensitivity of CLIC+LHC  collider to
anomalous $Wtb$ couplings at 95\% C.L. 
Only one of the couplings is assumed to deviate from the SM
at a time.
\label{tab6}}
\begin{center}
\begin{tabular}{lllcc}
$\sqrt{s_{ep}}$(TeV).&$\int{L}dt(fb^{-1})$ &$\delta^{sys}$
&$F_{2L}$ & $F_{2R}$ \\
\hline
 5 & 20 & 0     &-0.015, 0.016 & -0.038, 0.038   \\
 5 & 20 & 0.05  &-0.024, 0.029 & -0.044, 0.044   \\
 5 & 20 & 0.10     &-0.035, 0.303 & -0.051, 0.051   \\
 6.5 & 50 & 0  &-0.007, 0.008 & -0.026, 0.026   \\
 6.5 & 50 & 0.05     &-0.015, 0.017 & -0.034, 0.034   \\
 6.5 & 50 & 0.10  &-0.024, 0.290 & -0.041, 0.041   \\
\end{tabular}
\end{center}
\end{table}

\begin{figure}[htb]
  \begin{center}
  \epsfig{file=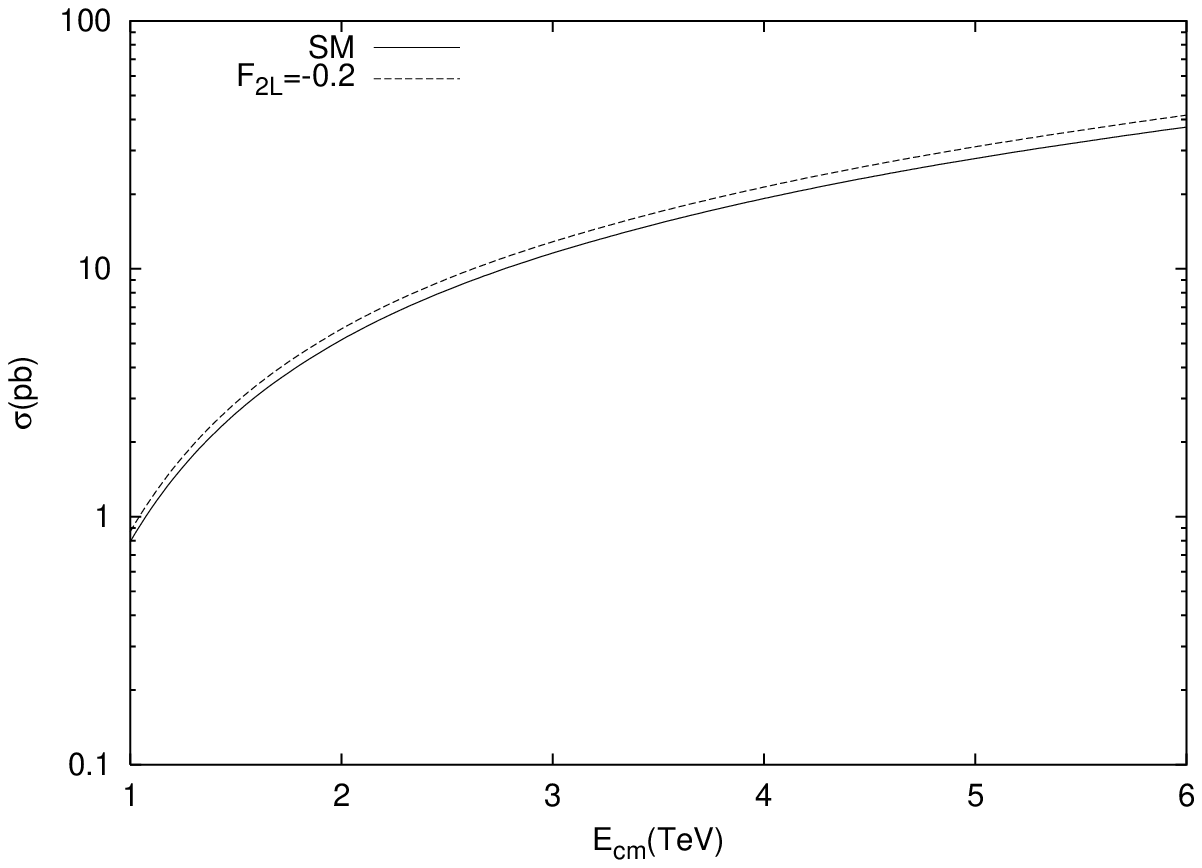}
  \end{center}         
 \ccap{fig1}{\footnotesize Energy dependence of the total cross
section for the single top quark production in the process 
$ep\rightarrow t\bar{\nu}+X$. }
\end{figure}                                                

\newpage

\begin{figure}[htb]
  \begin{center}
  \epsfig{file=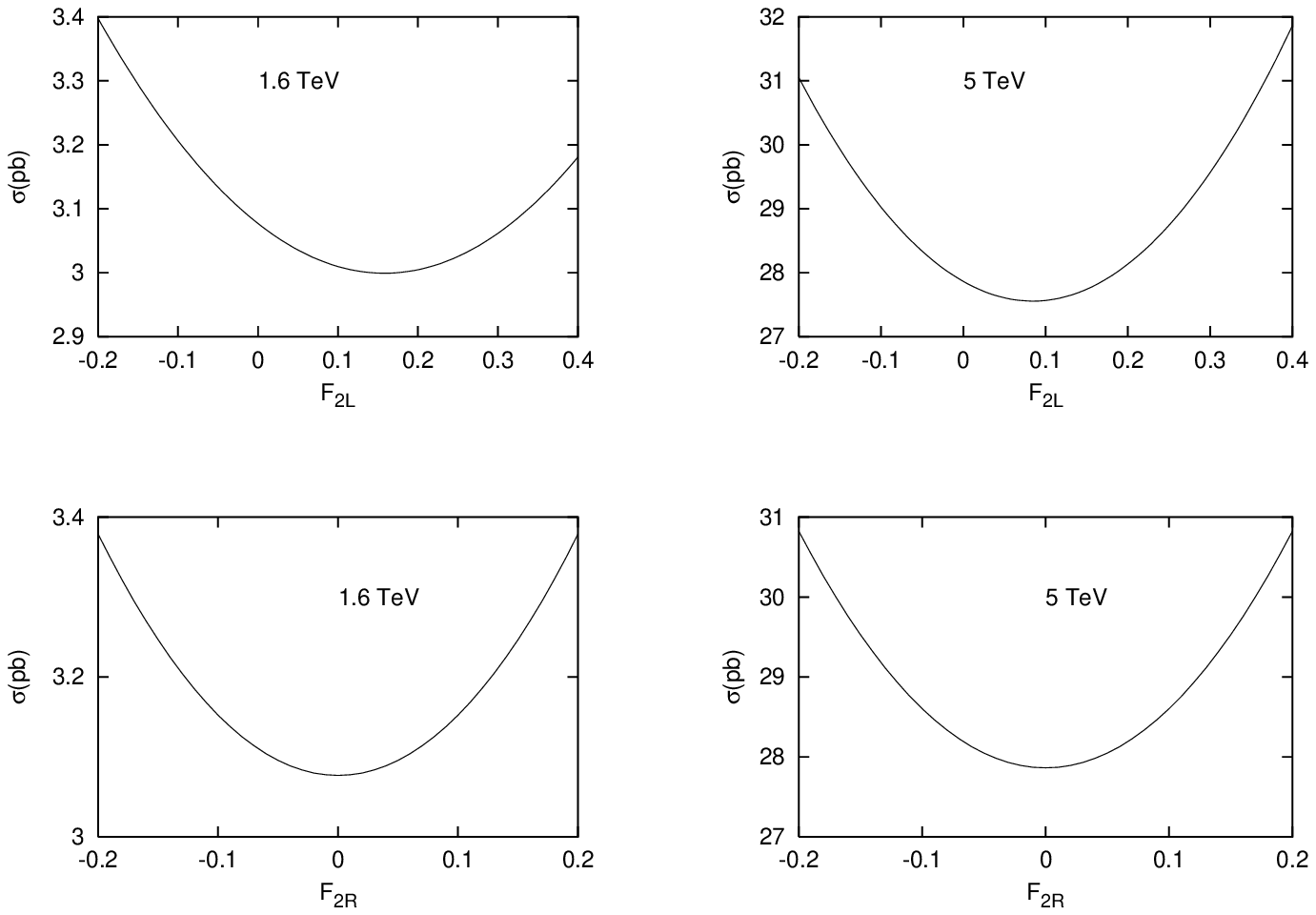}
  \end{center}
 \ccap{fig2}{\footnotesize Integrated   total cross sections 
as functions of anomalous couplings at possible TESLA+HERAp
(1.6 TeV) and CLIC+LHC (5 TeV) energies.}
\end{figure}

\begin{figure}[htb]
  \begin{center}
  \epsfig{file=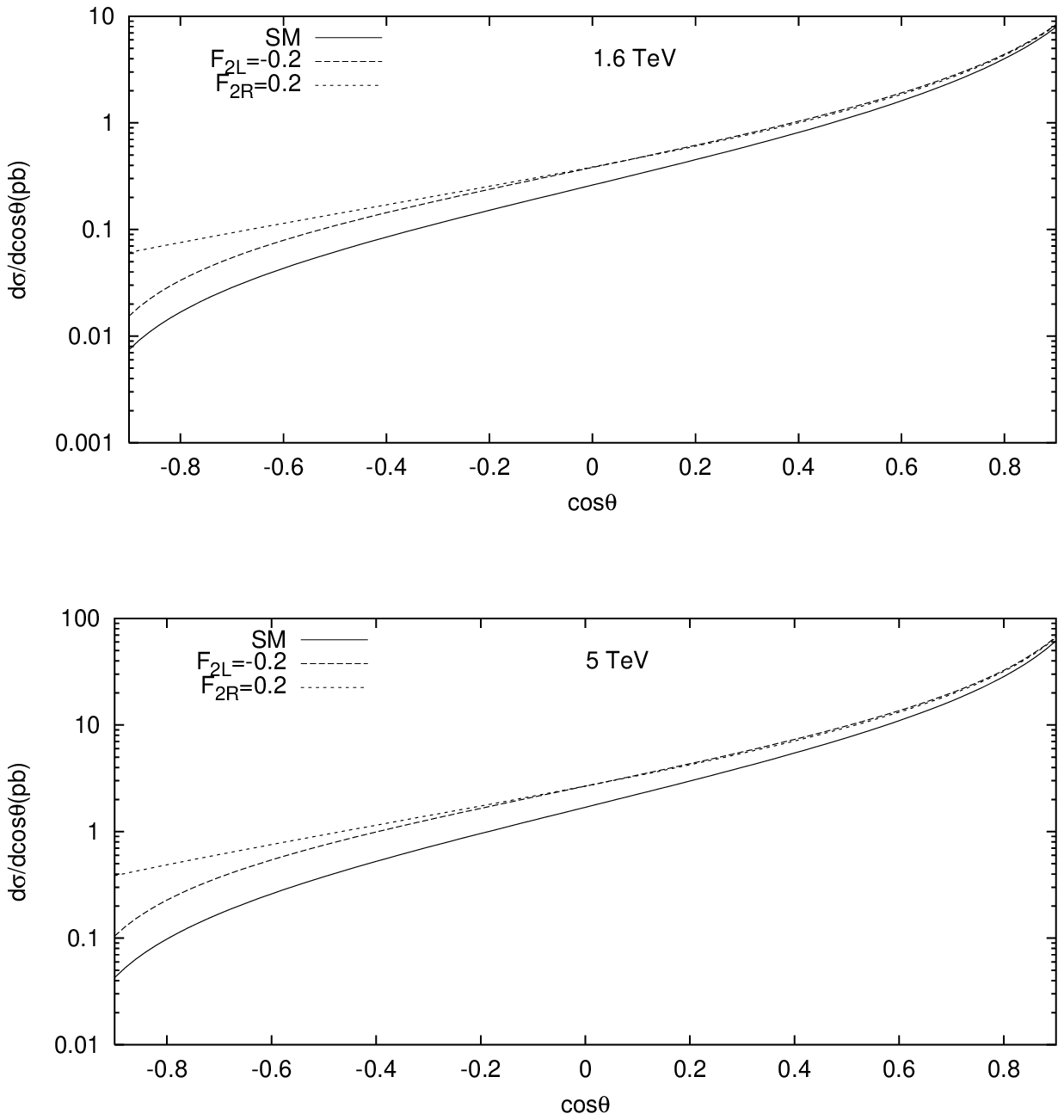} 
  \end{center}
 \ccap{fig3}{\footnotesize Angular distributions of the  top quark 
at TESLA+HERAp and CLIC+LHC energies in the center of mass 
frame of $t\bar{\nu}$. }
\end{figure}        

\begin{figure}[htb]
  \begin{center}
  \epsfig{file=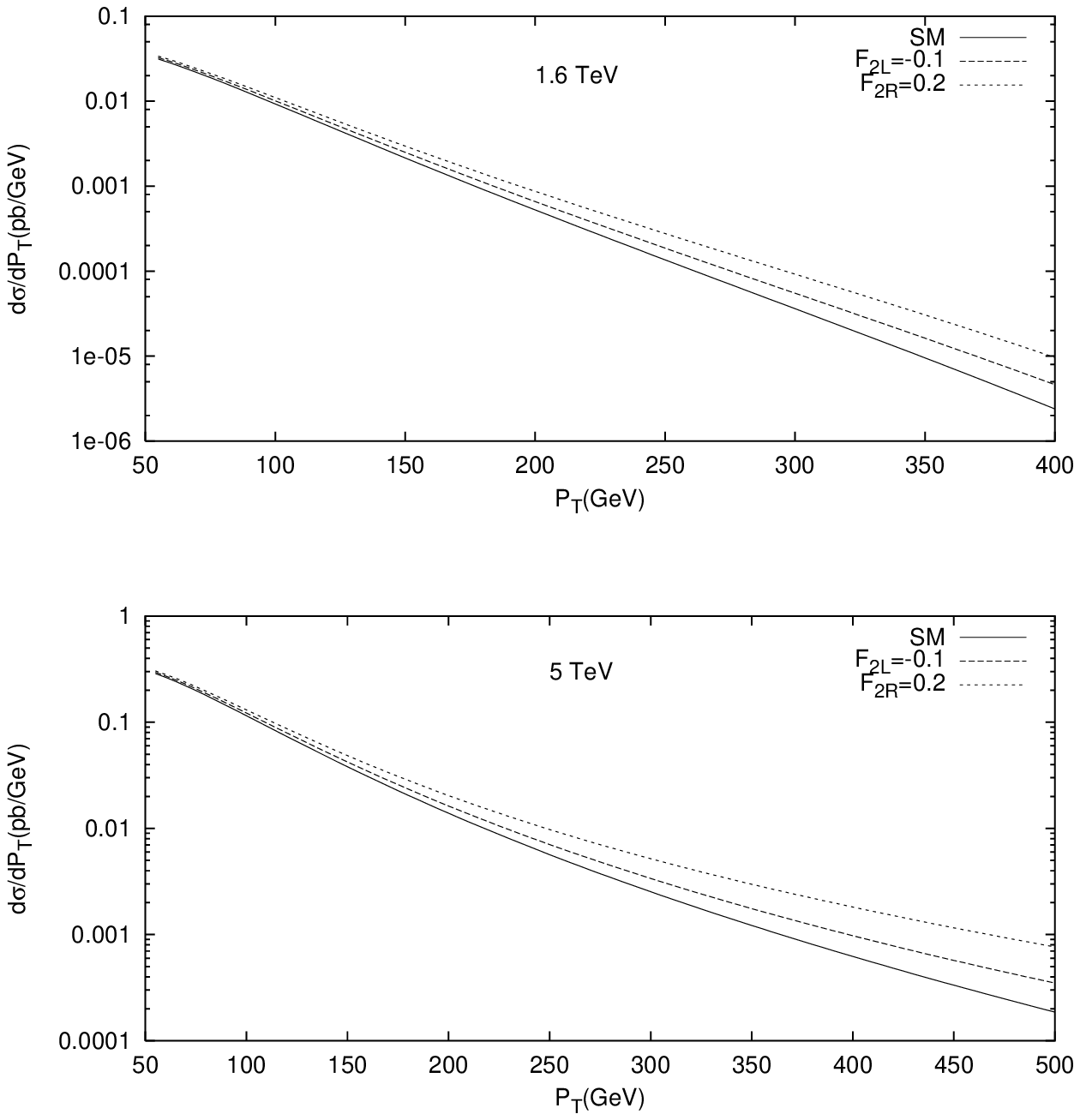}
  \end{center}
 \ccap{fig4}{\footnotesize $p_{T}$ distributions of the top quark at
TESLA+HERAp and CLIC+LHC energies. }
\end{figure}        

\end{document}